\documentclass[nofootinbib,letterpaper,aps,amsmath,amsfonts,12pt]{revtex4}
\usepackage[dvips]{graphicx}
\begin{document}

\setlength\arraycolsep{2pt} 
\def\Pcm#1{{\mathcal{#1}}}
\def\nn{\nonumber}
\def\er#1{eqn.\eqref{#1}}
 
\title{Non-Commutative Geometry and Twisted Conformal Symmetry}

\author{Peter~Matlock}
\email{pwm@imsc.res.in}
\affiliation{The Institute of Mathematical Sciences, Chennai, India}

\begin{abstract}
The twist-deformed conformal algebra is constructed as a Hopf algebra with
twisted co-product.  This allows for the definition of conformal symmetry
in a non-commutative background geometry.  The twisted
co-product is reviewed for the Poincar\'e algebra and the construction
is then extended to the full conformal algebra. It is demonstrated
that conformal invariance need not be viewed as incompatible with
non-commutative geometry; the non-commutativity of the coordinates
appears as a consequence of the twisting, as has been shown in the
literature in the case of the twisted Poincar\'e algebra.
\end{abstract}

\maketitle

\section{Introduction}
\label{intro}
Non-commutative geometry is often introduced by postulating that the
coordinates of space (or spacetime) do not commute, so that
\begin{equation}
[x^\mu,x^\nu]=i\theta^{\mu\nu}(x)
.\end{equation}
In order to formulate Physics in this context \cite{0106048,0109162},
fields must be defined on such a background, and of course they
inherit algebraic properties from those of the background itself. The
algebra of fields is used to construct an action and to express
physical principles.  One way of performing calculations is to avoid
dealing with the algebra of functions on non-commutative coordinates
as defined above, and instead formulate a theory in terms of the
algebra of functions on a commutative space, but with a deformed `star
product', defined such that these two algebras are isomorphic.
The application of such ideas and their appearance in String Theory
can be found in the well-known paper by Seiberg and Witten \cite{9908142}
and references therein.

In practice, dealing with an arbitrary non-commutativity function $\theta(x)$
is very difficult. Rather than attempt this, authors have mainly considered
constant, linear, and quadratic non-commutativity, referred to respectively 
as the `canonical structure', `Lie algebra' and `quantum space' 
types of non-commutative space, as explained in \cite{0001203}.
In the present paper, we consider the canonical 
structure case and will eventually show the property
\begin{equation}
\label{nccoors}
[x^\mu,x^\nu]=i\theta^{\mu\nu}
\end{equation}
where $\theta$ is a constant matrix, not depending on $x$.
At first glance, this relation seems to be more than just a
statement of non-commutativity; it appears to break 
Poincar\'e\footnote{It might be pointed out that \er{nccoors} only obviously breaks Lorentz 
invariance, as it is clearly invariant under $x\rightarrow x+a$.
This is only (na\"ively) true if the coordinate shift $a$ commutes with the coordinates, 
which is a non-trivial assumption in itself.}
invariance, and to introduce `preferred directions' into the formalism at
the outset. Given that such preferred directions disappear in the 
$\theta=0$ case, it is not clear that the simple limit $\theta\rightarrow 0$
reproduces the Lorentz symmetry which we are used to seeing in nature.
Furthermore, although the commutation relation \eqref{nccoors} has been
used as the basis of many analyses of physics in the language of quantum 
theory on non-commutative space, it does not make sense to talk of vectors 
or spin, which depend on representations of the Lorentz group, when it is 
not clear that there is a Lorentz group present.

These problems were pointed out and addressed systematically in the
interesting paper \cite{0409096} and in the lecture \cite{0408080}.\footnote{We
will follow more closely the notations of \cite{0409096}.} It is true
that the non-commutativity relation \eqref{nccoors} is not a
Poincar\'e-invariant statement, but it is twist-deformed Poincar\'e invariant.
`Relativistically invariant' in the usual sense
means that a theory is Poincar\'e-invariant, and this may be
consistently modified to mean invariance under twisted Poincar\'e
transformations in the case of a non-commutative background.  This is
demonstrated in \cite{0409096} and \cite{0408069} and we shall briefly
review the construction in order that the present work be 
somewhat self-contained. We note that the construction may also be 
generalised to the supersymmetric case, as investigated in \cite{0410164}.

The commutator \eqref{nccoors} does not look Poincar\'e-invariant, but
can be understood to be relativistically invariant in the twisted
Poincar\'e sense.  It also does not look conformally invariant, and
the purpose of the present paper is to show that it is consistent to
twist-deform the conformal algebra along the same lines as the
Poincar\'e algebra and therefore consider conformal symmetry in a
non-commutative background. The relation \eqref{nccoors} \emph{is}
twist-conformal invariant.

In section \ref{formalism} we review the construction of the Hopf algebra
and co-product used by the authors of \cite{0409096} to define the twisted 
Poincar\'e algebra. 
In section \ref{Tconfalg} we calculate the twisted co-product
for the generators of the full conformal algebra, including for 
completeness the Poincar\'e subalgebra.
Finally, in section \ref{noncomm} we give some explicit examples of
the twisted conformal transformations. As in the treatment in
\cite{0409096}, we show that the non-commutativity of \er{nccoors} is
a simple consequence of the deformed algebra, in this case the twisted
conformal algebra.
We conclude with some discussion in section \ref{ending}.
\section{Co-product and twist deformation}
\label{formalism}

Here we briefly explain the formalism so that we may apply it to the
conformal algebra; for a detailed treatment of Hopf algebras, the reader 
may consult the comprehensive reference \cite{QGguide}.

To deform the universal enveloping algebra $\Pcm{U}(A)$ of a Lie algebra $A$,
one first constructs a representation of $\Pcm{U}(A)$ in the tensor product 
$\Pcm{U}(A) \otimes \Pcm{U}(A)$ by defining a co-product
\begin{equation}
\Delta: \Pcm{U}(A)\rightarrow \Pcm{U}(A) \otimes \Pcm{U}(A)
.\end{equation}
Starting with the primitive co-product $\Delta_0$, defined by
\begin{equation}
\Delta_0(X) \equiv X \otimes 1 + 1 \otimes X
,\end{equation}
a twist element $\Pcm{F}\in\Pcm{U}(A) \otimes \Pcm{U}(A)$ may be chosen, 
and a twisted co-product $\Delta_t$ defined as
\begin{equation}
\label{twistcop}
\Delta_t(X) \equiv \Pcm{F} \Delta_0(X) \Pcm{F}^{-1}
.\end{equation}
The twist element must be an invertible element and satisfy the twist equation 
$\Pcm{F} (\Delta \otimes \textup{id}) \Pcm{F}=\Pcm{F} ( \textup{id}\otimes\Delta ) \Pcm{F}$
for the original co-product $\Delta=\Delta_0$ \cite{QGguide}. Consequently, the Hopf algebra
retains its properties and remains a Hopf algebra under the twist.
The twisted and untwisted co-products $\Delta_t$ and $\Delta_0$ of the generators satisfy 
the same Lie algebra; that is, the co-product map is an obvious homomorphism thereof.

In \cite{0409096} the algebra $A$ was taken to be the Poincar\'e
algebra $\Pcm{P}$, with momentum generators $P_\mu$ and Lorentz
generators $M_{\mu\nu}$. In that case, the twist element was taken to
be a so-called Abelian (since the $P_\mu$ form an Abelian subalgebra) 
twist,
\begin{equation}
\label{F}
\Pcm{F}=e^{\frac{i}2 \theta^{\mu\nu} P_\mu \otimes P_\nu}
,\end{equation}
constructed to reproduce the non-commutativity relation
\eqref{nccoors}, as we will see below.  Our main result will be that
this same twist element can be used to construct the twist-deformed conformal
algebra under which the non-commutativity \eqref{nccoors} is
invariant.  With this in mind, let us recall the full conformal
algebra for $d>2$, including the dilatation generator $D$, and the
special conformal transformation $K_\mu$,
\begin{eqnarray}
\label{confalg}
 {[ D , P_\mu ]} &=& i P_\mu \nn\\
 {[ D , K_\mu ]} &=& -i K_\mu \nn\\
 {[ K_\mu,P_\nu ]} &=& 2i(\eta_{\mu\nu}D - M_{\mu\nu} ) \nn\\
 {[ K_\rho , M_{\mu\nu} ]} &=& i( \eta_{\rho\mu}K_\nu -\eta_{\rho\nu}K_\mu ) \nn\\
 {[ P_\rho , M_{\mu\nu} ]} &=& i ( \eta_{\rho\mu} P_\nu - \eta_{\rho\nu}P_\mu ) \nn\\
 {[ M_{\mu\nu},M_{\rho\sigma} ]} &=& i ( \eta_{\nu\rho}M_{\mu\sigma} + \eta_{\mu\sigma}M_{\nu\rho} 
                                    - \eta_{\mu\rho}M_{\nu\sigma} - \eta_{\nu\sigma}M_{\mu\rho} ) 
.\end{eqnarray}
The conformal algebra contains the Poincar\'e algebra and the abelian
algebra of the $P_\mu$ as subalgebras. 
The represention of the
conformal algebra on an algebra $\Pcm{A}$ of functions on spacetime is
given by \cite{dFMS}
\begin{eqnarray}
\hat{P}_\mu &=& -i \partial_\mu \nn\\
\hat{M}_{\mu\nu} &=& i (x_\mu\partial_\nu-x_\nu\partial_\mu) \nn\\
\hat{D}&=&-i x\cdot\partial \nn\\
\hat{K}_\mu &=& 2x_\mu \hat{D} - x^2 \hat{P}_\mu 
,\end{eqnarray}
where $\hat{G}$ denotes the representation of $G$ on the algebra of functions $\Pcm{A}$.

In the primitive case, using the untwisted co-product $\Delta_0$, the multiplication 
in $\Pcm{A}$ is given by a map $m:\Pcm{A}\otimes\Pcm{A} \rightarrow \Pcm{A}$ defined 
simply as $m(f\otimes g) \equiv fg$. In the twisted case, this multiplication map must 
be modified by composition with the representation of the inverse twist element, to give 
the twisted map $m_t$ which is the `star product' in $\Pcm{A}$,
\begin{equation}
f \star g \equiv m_t(f\otimes g) = m \big( \Pcm{\hat{F}}^{-1} (f \otimes g) \big)
.\end{equation}
The algebra $\Pcm{A}$ endowed with the new twisted multiplication $m_t$ is then 
consistent with the twisted conformal algebra, exhibited explicitly in the following 
section. By `consistent' is meant that statements (i.e. physics) expressed in terms
of the elements and operations of the twisted algebra of functions are covariant under
the twist-deformed conformal algebra. This includes the relation \eqref{nccoors}, as we
show in section \ref{noncomm}.

\section{Twisted conformal algebra}
\label{Tconfalg}
Using the twisted co-product $\Delta_t$ given in \er{twistcop} with
the twist element of \er{F}, we may calculate the twisted co-products
of the generators. As mentioned in \cite{0408069}, the momentum
generators do not get twisted due to the commutativity of $P_\mu$;
\begin{equation}
\Delta_t(P_\mu)=P_\mu\otimes 1 + 1 \otimes P_\mu = \Delta_0(P_\mu)
.\end{equation}
For the Lorentz generators, $\Delta_t(M_{\mu\nu})$ has been calculated
in \cite{0408069} and \cite{0408080}, and is given by
\begin{equation}
\Delta_t(M_{\mu\nu}) = M_{\mu\nu}\otimes1 +  1\otimes M_{\mu\nu} 
 - \frac12 \theta^{\rho\sigma} \big(
     (\eta_{\mu\rho}P_\nu-\eta_{\nu\rho}P_\mu)\otimes P_\sigma
     +P_\rho \otimes (\eta_{\mu\sigma}P_\nu-\eta_{\nu\sigma}P_\mu)
   \big)
.\end{equation}
For the dilatation generator $D$ we find
\begin{equation}
\label{twistyD}
\Delta_t(D) = D\otimes1 +  1\otimes D
         + \theta^{\rho\sigma} P_\rho \otimes P_\sigma
,\end{equation}
while for the SCT generator $K_\mu$ a short calculation gives
\begin{eqnarray}
\label{twistyK}
\Delta_t(K_\mu) = K_\mu\otimes1 +  1\otimes K_\mu
         &+& \theta^{\rho\sigma} \big(
                     (\eta_{\rho\mu}D + M_{\rho\mu})\otimes P_\sigma
                     + P_\rho \otimes (\eta_{\sigma\mu}D + M_{\sigma\mu})
                     \big) \nn\\
         &+&\frac14 \theta^{\rho\sigma}\theta^{\lambda\pi}\big(
           (  \eta_{\lambda\mu}P_\rho -\eta_{\lambda\rho}P_\mu + \eta_{\rho\mu}P_\lambda)
                        \otimes P_\pi P_\sigma \nn\\
         & &{} + P_\lambda P_\rho \otimes
           ( \eta_{\pi\mu}P_\sigma - \eta_{\pi\sigma}P_\mu + \eta_{\sigma\mu}P_\pi)
        \big)
.\end{eqnarray}
As a consistency check, it may be shown explicitly that these objects satisfy
 the conformal algebra \eqref{confalg}, as by construction they should.

\section{Non-commutativity and transformations}
\label{noncomm}
The non-commutativity relation \eqref{nccoors} is a consequence of the 
twisted algebra, as shown in \cite{0409096}. Evaluating the commutator of
 $x_\mu$ with $x_\nu$ in $\Pcm{A}$, we have
\begin{equation}
 [ x_\mu , x_\nu ]_{\Pcm{A}} \equiv m_t( x_\mu \otimes x_\nu ) - m_t( x_\nu \otimes x_\mu ) = \theta_{\mu\nu}
.\end{equation}
In particular, $\theta_{\mu\nu}$ is an invariant under twisted Poincar\'e transformations. 
Significanty, $\theta_{\mu\nu}$ is also invariant under twisted conformal transformations, 
as we will show at the end of this section.

Now, let us choose some simple example functions and demonstrate explicitly
that they transform correctly under the twisted conformal symmetry.
Following \cite{0409096} consider the second-rank tensor
$f_{\rho\sigma}=x_\rho x_\sigma$.  In that paper, the action of the
Lorentz algebra on this object is calculated to show how the
twisted covariance works in practice on a Lorentz tensor.  A similar
test can be performed using the twisted conformal algebra.  The
actions of the original (untwisted) dilatation generator $D$ and SCT
generator $K_\mu$ on $f_{\rho\sigma}$ are given by
\begin{equation}
\label{utD}
D f_{\rho\sigma} = -2 i f_{\rho\sigma} 
,\end{equation}
showing that $f_{\rho\sigma}$ has conformal dimension two, and
\begin{equation}
\label{utK}
K_\mu f_{\rho\sigma} = -4i x_\mu x_\rho x_\sigma + i x^2 (\eta_{\mu\rho}x_\sigma + \eta_{\mu\sigma}x_\rho)
.\end{equation}

In the twisted case, $f_{\rho\sigma}$ is replaced by the twisted
object $f^t_{\rho\sigma}=m_t(x_{\rho} \otimes x_{\sigma})$, 
and the conformal generators are now applied
through the twisted co-product, so that for a generator $G$,
\begin{equation}
G^t f^t_{\rho\sigma} = m_t \big( \Delta_t(\hat{G}) (x_\rho \otimes x_\sigma ) \big)
.\end{equation}
Using the twisted co-product from \er{twistyD}, we first calculate
\begin{equation}
\Delta_t(\hat{D}) x_\rho \otimes x_\sigma = -2i x_\rho \otimes x_\sigma - \theta_{\rho\sigma} 1 \otimes 1
\end{equation}
so that, using an expansion for $\Pcm{F}^{-1}$ (which terminates in this case),
\begin{equation}
\Pcm{\hat{F}}^{-1} \Delta_t(\hat{D}) x_\rho \otimes x_\sigma 
 = -2i x_\rho \otimes x_\sigma 
.\end{equation}
We can now read off
\begin{equation}
\label{twistedDonf}
D^t f^t_{\rho\sigma} = -2i f^t_{\rho\sigma}
,\end{equation}
analogous to the untwisted case, \er{utD}, showing that the conformal dimension of the 
twisted tensor $f^t$ is two, under the twisted conformal algebra.
Now, considering the action of the twisted co-product of $K_\mu$ from \er{twistyK} on $f^t$, we have
\begin{eqnarray}
\Delta_t(\hat{K}_\mu) x_\rho \otimes x_\sigma &=&
 -i (2x_\mu x_\rho - x^2 \eta_{\mu\rho}) \otimes x_\sigma
 -i x_\rho \otimes (2x_\mu x_\sigma - x^2 \eta_{\mu\sigma}) \nn\\
&&{}+\theta^{\lambda\tau}\bigg(
                    (x_\lambda\eta_{\mu\rho} -x_\rho\eta_{\lambda\mu} - x_\mu\eta_{\lambda\rho} )
                     \otimes \eta_{\tau\sigma} \nn\\
                  &&\qquad\qquad+\eta_{\lambda\rho} \otimes 
		     (x_\tau\eta_{\mu\sigma} -x_\sigma\eta_{\tau\mu} - x_\mu\eta_{\tau\sigma} )
                    \bigg) 
.\end{eqnarray}
Acting on this with $\Pcm{F}^{-1}$ we find
\begin{equation}
\Pcm{\hat{F}}^{-1} \Delta_t(\hat{K}_\mu) x_\rho \otimes x_\sigma
=  -i (2x_\mu x_\rho - x^2 \eta_{\mu\rho}) \otimes x_\sigma
   -i x_\rho \otimes (2x_\mu x_\sigma - x^2 \eta_{\mu\sigma})
,\end{equation}
and performing the multiplication $m$ we obtain
\begin{equation}
\label{twistedKonf}
K_\mu^t f^t_{\rho\sigma} = 
-4i x_\mu x_\rho x_\sigma + i x^2 (x_\sigma\eta_{\mu\rho} + x_\rho\eta_{\mu\sigma})
.\end{equation}
This equation is in agreement with the untwisted case \eqref{utK},
reflecting that the full conformal algebra is represented.

Now, returning to the question of invariance of the non-commutativity matrix 
$\theta_{\rho\sigma}$ under the twisted conformal group, we may 
write $i\theta_{\rho\sigma}$ as the commutator $[x_\rho,x_\sigma]$ in the twisted algebra
as we did at the beginning of this section. This is nothing but the antisymmetric 
part of the example tensor $f^t_{\rho\sigma}$ which we have considered above.
Noticing that the right-hand sides of equations \eqref{twistedDonf} and 
\eqref{twistedKonf} are symmetric in $\rho$ and $\sigma$, we immediately have
\begin{equation}
D^t \theta_{\rho\sigma} = 0
\qquad \textup{and} \qquad 
K_\mu^t \theta_{\rho\sigma} = 0
,\end{equation}
showing the expected result that $\theta_{\rho\sigma}$ is invariant under 
twisted conformal symmetry.

\section{Final remarks}
\label{ending}
We have constructed the twist-deformed conformal algebra as a Hopf
algebra with a twisted co-product. The twist element $\Pcm{F}$ has
been chosen so that the resulting algebra of functions reproduces the
non-commutativity of coordinates often considered to define
non-commutative geometry.  The conclusion is that a theory formulated
in the twisted algebra of functions, by using the star-product in
place of the traditional product in the Lagrangian, will be invariant
under the twist-deformed conformal algebra.

It would be interesting to investigate the full conformal group and
examine global and local transformations. This could be done along the
lines of the twisted diffeomorphism invariance constructed in
\cite{0411224}.  Using this structure in field theory it should be
possible to construct a well-defined notion of Conformal Field Theory
in a non-commutative background.  A natural extension would be to
investigate the two-dimensional case, where the conformal algebra is
infinite-dimensional.

\section*{Acknowledgements}
The author is grateful to T.~R.~Govindarajan for interesting
discussions on non-commutative and fuzzy geometry.


\end{document}